\documentclass[preprint,aps]{revtex4}
\newcommand{\be}{\begin{equation}}
\newcommand{\ee}{\end{equation}}
\newcommand{\bea}{\begin{eqnarray}}
\newcommand{\eea}{\end{eqnarray}}
\newcommand{\bref}[1]{(\ref{#1})}
\newcommand{\nn}{\nonumber}
\newcommand{\A}{\alpha}  
 \newcommand{\D}{\delta} 
\newcommand{\ep}{\epsilon}

           \newcommand{\s}{\sigma}
          \newcommand{\w}{\omega}
          
\newcommand{\h}{\eta}           
           
\newcommand{\W}{\Omega}

\def\6{\partial}\def\7{\tilde}
\def\8{\widehat}  
\def\vs{\vskip 3mm}
\def\pa{\partial}

\def\CL{{\cal L}}

\def\CH{{\cal H}}
\def\CM{{\cal M}}

\def\l{{\ell}}

\def\MC{Maurer-Cartan }
\def\NR{non-relativistic }
\def\too{{\;\to\;}}
\def\={{\;=\;}}\def\+{{\;+\;}}\def\-{{\;-\;}}
\def\we{\wedge}
\begin{document}

\preprint{CECS-PHY-06/02, UB-ECM-PF-06/06, Toho-CP-0679}

\title{Newton-Hooke Algebras, Non-relativistic
Branes and Generalized pp-wave Metrics}

\author{Jan Brugu\'es}
\affiliation{%
Departament ECM, Facultat de F{\'\i}sica, Universitat de Barcelona,
%\\CER for Astrophysics, Particle Physics and Cosmology,
Diagonal      647, E-08028     Barcelona,  Spain
}%
\author{Joaquim Gomis}%
%\email{gomis@ecm.ub.es} %happens error when included
\affiliation{
Departament   ECM,   Facultat   de F{\'\i}sica, Universitat de Barcelona,
CER for Astrophysics, Particle Physics and Cosmology,
Diagonal      647, E-08028     Barcelona,  Spain\\
Centro de Estudios Cient\'{\i}ficos (CECS)
Casilla 1469 \\ Valdivia, Chile}%
\author{ Kiyoshi Kamimura}
%\email{kamimura@ph.sci.toho-u.ac.jp}%happens error when included
\affiliation{Department of Physics, Toho University
Funabashi  274-8510,  Japan }%

\date{March 3, 2006, rev. April 7, 2006}% It is always \today, today,

\begin{abstract}
{The Newton-Hooke algebras in $d$ dimensions 
are constructed as contractions of dS(AdS) algebras.
Non-relativistic brane actions are WZ terms of these Newton-Hooke
algebras. The NH algebras appear also as subalgebras of
multi-temporal relativistic conformal algebras, $SO(d+1, p+2)$. We
construct generalizations of pp-wave metrics from these algebras.
 }
\end{abstract}

%\pacs{Valid PACS appear here}
% PACS, the Physics and Astronomy Classification Scheme.
% 11.10.Ef 	Lagrangian and Hamiltonian approach
% 11.25.-w 	Strings and branes
% 03.65.Fd 	Algebraic methods 
% 45.20.D- 	Newtonian mechanics

\maketitle

\section{Introduction}
In order to understand the AdS/CFT correspondence at string level one consider
sectors of string theory in which certain degrees of freedom decouple from the
rest of degrees of freedom in suitable limits.
Such decoupling sectors have different asymptotic symmetries
compared to the full theory.
The best known example is the BMN \cite{Berenstein:2002jq} sector of the
string theory in $AdS_5\times S^5$. The relevant symmetry of the BMN sector
is a super pp-wave algebra, which is a particular contraction
of the $ su(2,2|4)$, % symmetry of $AdS_5\times S^5$,
see for example \cite{Hatsuda:2002xp}.

Non-relativistic string theory  in flat space \cite{Gomis:2000bd},
see also \cite{Danielsson:2000gi,Garcia:2002fa,Brugues:2004an} is another 
example of a 
consistent decoupled sector of the bosonic string theory. The basic idea
of the decoupling limit is to take a
particular non-relativistic limit such that only states satisfying a Galilean
invariant dispersion relation have finite energy, while the rest decouple.
This  limit has been extended to other  relativistic
supersymmetric string theories \cite{Gomis:2004pw,Gomis:2005pg}.
In particular for the case of $AdS_5\times S^5$ we have found the 
non-relativistic string  theory in a suitable gauge
reduces to a supersymmetric free field theory in $AdS_2$.

%Here we  will construct  \NR bosonic algebras as brane contractions
%of $AdS$ ($dS$) algebras. 
{Here we  will construct  \NR bosonic algebras by contracting
$AdS$ ($dS$) algebras.} 
They are generalization of the
Newton-Hooke (NH) algebras { \cite{BacryLL, Derome}}
to the brane systems. We will apply the 
non-linear realization {to} these algebras and we will see how it is 
possible to construct the non-relativistic brane actions as WZ terms. This
property is due to the fact that one can construct a closed and
invariant $p+2$ form but the associated $p+1$ form
cannot be written in terms of left invariant forms. The
$p+2$ Eilenberg-Chevalley cohomology is non-trivial.

We will also see how the NH algebra can be extended with
non-central charges. It is due to an existence of non-trivial
vector valued 2 form on a representation of the stability group.

We will also construct generalization of pp-wave  metrics
from these extended algebras. For the case of the ordinary NH algebra 
we construct the pp-wave metric \cite{Gibbons:2003rv}
 and its generalization with angular momentum, the gyraton
 \cite{Frolov:2005zq}. For the NH brane algebras we will see these metrics
live in a space with more than one time-like directions.
Related to this point we will also show that the
brane NH algebras in $d$ dimensions are subalgebras
of multi-temporal conformal algebras, $SO(d+1,p+3)$.

The organization of the paper is as follows. In section II
we study the brane NH algebras. In section III we construct the
actions as WZ terms. In section IV we construct the non-central
extended NH algebras. In section V we construct the generalized pp-wave
metrics {from} these algebras and finally in section VI we see the algebras
are  subalgebras of multitemporal conformal algebras.

%%%%%%%%%%%%%%%%%%%%%%%%%%%%%%%%%%%%%%%%%%%%%%%%%%%%%%%%%%%%%%%%%%%%%%%%%%%%
\section{Newton-Hooke Algebras}

The particle Newton-Hooke algebra %{ \cite{NH} }
NH$_{+(-)}$ in $d$ dimensions is
obtained by contraction of dS(AdS) algebra in a similar manner as
the Galilei algebra is obtained by contraction of Poincare algebra
\cite{BacryLL}. The  dS(AdS) algebra is
\bea
\label{dsadsal}
 \left[M_{mn},M_{rs}\right]&=&i(\eta_{mr}M_{ns}+\eta_{ns}M_{mr}
 -\eta_{ms}M_{nr}
  -\eta_{nr}M_{ms})
\nonumber \\
  \left[M_{mn},P_{r}\right]&=&i(\eta_{mr}P_{n}-\eta_{nr}P_{m})
\nn\\
  \left[P_{m},P_{n}\right]&=& \pm\frac{i}{R^{2}}M_{mn} \ ,
\eea
where $m=0,1,...,d-1$, $\eta_{mn}=(-;+...+)$, $\pm$ in the last
commutator corresponds to dS(AdS) and  $R$ is the radius of the
dS(AdS) space. It is the isometry group of the homogeneous space with a
constant curvature $\pm R^{-2}$.
It is embedded in a $d+1$ dimensional space of
coordinates $u^M, (M=0,1,...,d-1,d) $ with a constraint
\be
\eta_{mn}u^mu^n+\eta_{dd} (u^d)^2=\eta_{dd} R^2 ,
\ee
where $\eta_{MN}=(-;+...+,\eta_{dd})$ and $\eta_{dd}=+1$
for the dS and  $\eta_{dd}=-1$ for AdS space.

In this paper we study the Newton-Hooke algebras associated {with}
a class of  non-relativistic $p+1$ dimensional
objects (p-branes) living in the $d$ dimensional space-time,
that we call NH brane algebras  and  are also denoted by $NH_{\pm}$. These
algebras are obtained by a suitable contraction of the $d$ dimensional
dS(AdS) algebra. 
The contraction is done by performing the following scaling on the generators
\bea
P_\mu&=&~\frac{1}{\w}\tilde{P}_\mu,~~~~~~~~~~\mu=0,1,...,p,
\nn\\
P_a&=&\tilde{P}_a,~~~~~~~~~~~~~~a=p+1,...,d-1,
\nn\\
M_{\mu\nu}&=&~\tilde{M}_{\mu\nu}
\nn\\
M_{\mu b}&=&\w\tilde{M}_{\mu b}
\nn\\
M_{ab}&=&\tilde{M}_{ab}
\nn\\
R&=&\w\;\7R,
\label{contr}
\eea
where $\w$ is a dimensionless parameters, and we  take $\w\to\infty$.
The greek indices denote the longitudinal coordinates $(\mu=\{0,...,p\})$
to the brane
whereas latin indices are the transverse coordinates $(a=\{p+1,..., d-1
\})$ to the brane.
This scaling is suggested by the non-relativistic limit of
relativistic branes \cite{Gomis:2000bd,Danielsson:2000gi,Garcia:2002fa,
Brugues:2004an,Gomis:2004pw,Gomis:2005pg}.
We get for  the non-zero commutators\footnote{We will omit "tilde"on the 
rescaled generators hereafter.}
\bea
\left[P_\mu,P_\nu\right]&=&\pm
\frac{i}{R^2} M_{\mu\nu} \ , \qquad
\left[M_{\mu\nu},P_\rho\right]=i(\eta_{\mu\rho}P_\nu -
\eta_{\nu\rho}P_\mu)  \ ,
\nonumber\\
\left[M_{\mu\nu},M_{\rho\s}\right]&=&
i(\eta_{\mu\rho}M_{\nu\s}+\eta_{\nu\s}M_{\mu\rho}
-\eta_{\mu\s}M_{\nu\rho}
-\eta_{\nu\rho}M_{\mu\s})\ , %\nonumber
\label{contra1}\\
\left[M_{ab},P_c\right]&=&i(\eta_{ac}P_b - \eta_{bc}P_a) \ ,
\nonumber\\
\left[M_{ab},M_{cd}\right]&=&
i(\eta_{ac}M_{bd}+\eta_{bd}M_{ac}
 -\eta_{ad}M_{bc}-\eta_{bc}M_{ad}),
\label{contra3}\\
\left[P_\mu,P_a\right] &=&\pm \frac{i}{R^2} M_{\mu a}\ ,
\qquad 
\left[M_{\mu a},P_\nu\right] =i\eta_{\mu \nu}P_a \ ,
\nonumber\\
\left[M_{\mu\nu},M_{\rho c}\right]&=&i
(\eta_{\mu\rho}M_{\nu c}-\eta_{\nu\rho}M_{\mu c})\ ,\quad
\left[M_{\mu a},M_{bc}\right]=-i
(\eta_{ab}M_{\mu c}-\eta_{ac}M_{\mu b}).
\label{contra2}
\eea
The generators $P_\mu$ and $M_{\mu\nu}$ form a $p+1$ dimensional dS(AdS)
subalgebra \bref{contra1} while the $P_a$ and $M_{ab}$ form a 
$d-p-1$ dimensional Euclidean subalgebra \bref{contra3}.
If we further take the limit $ R\,\rightarrow\infty$ we
obtain the non-relativistic Galilean brane algebras \cite{Brugues:2004an}.

%%%%%%%%%%%%%%%%%%%%%%%%%%%%%%%%%%%%%%%%%%%%%%%%%%%%%%%%%%%%%%%%%%

\section{NH Brane Actions from Non-linear Realizations}\label{sectBrane}

Let us consider the coset $NH_\pm/H$ with the stability group
$H=\{M_{\mu\nu},M_{ab}\}$. We parametrize the coset element as
\bea
\label{p2} g&=&g_{0}\;e^{i y^{a}P_{a}}e^{iv^{\mu a}M_{\mu a}}.
\eea
$g_{0}$ is an element of dS$_{p+1}$(AdS$_{p+1})$
and the corresponding Maurer-Cartan one form is
\be 
\W_0\=-i g_0^{-1}dg_0 =e^\mu
P_\mu+\frac12 w^{\mu\nu}M_{\mu\nu}, 
\ee
where $e^\mu$ and $w^{\mu\nu}$ are dS$_{p+1}$(AdS$_{p+1})$ vielbein and spin
connection, the explicit forms depend on the parametrization of $g_0$.
The Maurer-Cartan one form $\W_0$ verifies the   Maurer-Cartan equation
$d\W_0+i\W_0\we\W_0=0$. In components it is
\be 
de^\mu+{w^\mu}_{\nu}e^\nu=0,\qquad
dw^{\mu\nu}+{w^\mu}_{\rho}w^{\rho\nu}=\pm\frac{1}{R^2}e^\mu e^\nu.
\ee
The second equation shows that the longitudinal space has
constant curvature $\pm R^{-2}$.

The total Maurer-Cartan one form, $\Omega$ is
\bea
\Omega=
-ig^{-1}dg=L^\mu P_\mu+L^a{P_{a}}+L^{\mu a}M_{\mu a} +
\frac12
L^{\mu\nu}M_{\mu\nu}+\frac12 L^{ab}M_{ab}
\eea 
where \footnote{ 
Recall that here and below in the text when $\pm$ or $\mp$ appear,
the upper indices correspond to dS and the lower indices to AdS.
}
\bea 
L^\mu&=&e^\mu,
\nn\\
L^a&=&dy^a+e^\mu {v_{\mu}}^a,
\nn\\
L^{\mu \nu}&=&w^{\mu\nu},
\nn\\
L^{\mu a}&=& dv^{\mu a} +w^{\mu\nu}{v_\nu}^a \mp
 e^\mu\frac{y^a}{R^2},
\nn\\
L^{ab}&=&0.
\label{W1}
\eea
These left invariant forms verify the Maurer-Cartan equation $d\W+i\W\we\W=0$, 
more explicitly,
 \bea
dL^\mu&+&L^\rho{L_\rho}^\mu=0,
\nn\\
dL^a&+&L^\rho{L_\rho}^a+L^c{L_c}^a=0,
\nn\\
dL^{\mu \nu}&+&L^{\mu
\rho}{L_\rho}^\nu\mp\frac{1}{R^2}L^{\mu}{L}^\nu=0,
\nn\\
dL^{\mu a}&+&L^{\mu \rho}{L_\rho}^a+L^{\mu c}{L_c}^a
\mp\frac{1}{R^2}L^{\mu}{L}^a=0,
\nn\\
dL^{ab}&+&L^{ac}{L_c}^b=0.
\label{MCp} 
\eea

We can construct the non-relativistic Newton-Hooke brane action as
a WZ term of the NH algebra. There is a closed $p+2$ form
invariant under $H$ using the MC one forms $L$'s in 
\bref{W1}\footnote{Our convention is $\ep^{01,...,p}=-\ep_{01,...,p}=+1$.}
 \be
\Omega_{p+2}=\frac{(-1)^{p+1}}{p!}
\ep_{\mu_0\mu_1,...,\mu_p}L^{\mu_0}...L^{\mu_{p-1}}L^{\mu_pa}L_a.
\label{Wp+2}
\ee 
The $H$ invariance is manifest and the closure is
shown by using the Maurer-Cartan equations \bref{MCp}.
There exists a $p+1$ form $\Omega_{p+1}$ such that 
$\Omega_{p+2}=d\Omega_{p+1}$,
\bea
\Omega_{p+1}&=&-\frac{\ep_{\mu_0\mu_1,...,\mu_p}}{p!}e^{\mu_0}...e^{\mu_{p-1}}
\left[{v^{\mu_p}}_a dy^a+ e^{\mu_p}\left(\frac{v_{\nu a}v^{\nu
a}}{2(p+1)}\pm \frac{y_{a}y^{a}}{2R^2}\right)\right]. \label{Wp+1}
\eea 
The $p+1$ form (\ref{Wp+1}) is not expressible in terms of the
left invariant one forms $L$'s in \bref{W1}. Therefore we have a non-trivial
$p+2$ cohomology group in the sense of Eilenberg-Chevalley.

By taking the pullback of the $p+1$ form  \bref{Wp+1} on the world sheet;
\be
S_{WZ}\equiv T \int \Omega_{p+1}^{*} \=T\int d^{p+1}\xi\;\CL ,
\ee
where $T$ is the non relativistic $p$-brane tension and $\xi^j(j=0,1,...,p)$ 
are the parameters of the worldvolume. $\Omega_{p+1}^*$ is given by
\bea
\label{2-}
\Omega_{p+1}^*&=&
e \left[{v^\mu}_a {e_\mu}^j\pa_{j}y^a+
\left(\frac{v_{\nu a}v^{\nu a}}{2}\pm
\frac{(p+1)}{2R^2}y_{a}y^{a}\right)\right]d^{p+1}\xi\equiv \CL \;d^{p+1}\xi,
\eea
where
$ {e_\mu}^j$ is the inverse of the $p+1$-bein ${e_j}^\mu$ and
$e=\det{ ({e_j}^\mu)}$.
Since $v^{\mu a}$ are non-dynamical Goldstone fields
we can express them  using their equations of motion as
\be
v_{\mu a}\+{e_\mu}^j\pa_{j}y_a\=0
\label{eomv}\ee
which are  equivalent to impose $H$ invariant constraints
\be
L^a=0,
\ee
in other words the inverse Higgs mechanism \cite{Ivanov:1975zq}.

In terms of the dynamical variables $y^{a}$ and $x^{\mu}$
the Lagrangian becomes
\be
\CL\=\frac{\sqrt{-g}}{2} \left[-g^{ij}\pa_{i}y^a\pa_{j}y_a\pm
\frac{(p+1)}{R^2}y_{a}y^{a}\right],
\label{Lagnrp}
\ee
where $g_{ij}$ is the induced metric of the longitudinal space
dS$_{p+1}$(AdS$_{p+1}$)
\be
g_{ij}=\h_{\mu\nu}{e_i}^\mu{e_j}^\nu=\7g_{\mu\nu}(x)\pa_ix^\mu \pa_jx^\nu,
\quad g^{ij}g_{jk}={\D^i}_k,\quad
g =\det g_{ij}=-e^2.
\ee
In the static gauge $x^j=\xi^j$ the Lagrangian describes a set of free scalar 
fields $y^a$ of mass$^2=\mp\frac{p+1}{R^2}$ in the dS$_{p+1}$(AdS$_{p+1}$) 
space with the metric $\7g_{\mu\nu}(x)$.

In the limit $R\to \infty$ we recover the Galilean result \cite{Brugues:2004an}
 since in that case the dS(AdS) algebra \bref{dsadsal}
goes to the Poincare algebra and the Newton-Hooke algebra
(\ref{contra1} - \ref{contra2}) becomes the Galilean algebra.

The Lagrangian \bref{Lagnrp} obtained here as a WZ term can also be
obtained from the relativistic brane action in the dS(AdS)\footnote{
Actions of branes in AdS  have constructed also using 
the non-linear realization methods in \cite{Delduc:2001tb}. 
}
by
a \NR limit. In  this limit the coordinates are rescaled as
\bea 
X^{\mu}&=& \w x^{\mu} \ , \qquad X^a= x^a \ ,\  \qquad
R = \w \7R \ , \  \qquad T = \w^{1-p} \7T.
\label{reesc}
\eea
In the limit $\w\to\infty$ of the Lagrangian
there appears a divergent surface term but can be
compensated by the presence of the B-field \cite{Gomis:2000bd}.
The resulting finite Lagrangian becomes one in  \bref{Lagnrp} .

%%%%%%%%%%%%%%%%%%%%%%%%%%%%%%%%%%%%%%%%%%%%%%%%%%%%%%%%%%%%%%%%%%

\section{Non-central Extensions of Newton-Hooke algebras}\label{sectcohomology}

Here we want to study the extensions of the NH algebra \bref{contra1}-
\bref{contra2}. We look for closed 2-forms constructed from the left 
invariant one forms
\bref{W1} that transforms non-trivially under the stability group $H$,
in particular under $M_{\mu\nu}$\footnote{ For ordinary supersymmetry
the role of Lorentz one-forms  was considered in \cite{Aldaya:1984gt}.}.
As we will see shortly these forms will be responsible for the
existence of non-central extensions of the NH algebra.

Let us consider a vector valued 2-form $L^c {L_c}^\mu$ \footnote{
Other tensor valued two form like $L^\rho {L_\rho}^a$ is  exact
and can be written in terms of invariant forms,
$L^\rho {L_\rho}^a\sim dL^a$.}. This form is  not  closed but we have
\be
\label{v2f}
d (L^c {L_c}^\mu)\sim -L^\rho{L_\rho}^c{L_c}^\mu.
\ee
where $\sim$ means up to terms depending on $L^{\mu\nu}, L^{ab}$, 
in other words $L^{\mu\nu}\sim L^{ab}\sim0$.
If we introduce a new tensor valued one form $L^{\mu\nu}_z$ verifying
\be 
d L^{\mu\nu}_z\+L^{\mu c}{L_c}^\nu \sim 0
\label{mct1f}
\ee
we can see that the 2-form
$L^c {L_c}^\mu+ L^\rho{{L_z}_\rho}^\mu$
is closed. This condition
indicates  locally the existence of a  1-form $L^{\mu}_z$
such that
\be
dL^{\mu}_z+L^c {L_c}^\mu+ L^\rho{{L_z}_\rho}^\mu\sim 0.
\label{mcv1f}
\ee
Now we take into account the tensor properties of ${L_z}^{\mu\nu}$
and $L^{\mu}_z$ under $M_{\mu\nu}$. % in \bref{MCp}, t
The  equations \bref{mct1f} and \bref{mcv1f} become
\bea
dL^{\mu}_z&+&{L^\rho}{{L_z}_\rho}^\mu+{L^c}{{L_c}}^\mu+{L^{\mu\rho}{L_z}_\rho}
=0,
\label{MCp0}
\\
dL^{\mu\nu}_z&+&{L^{\mu c}}{L_c}^{\nu}
+L^{\mu\rho}{{L_z}_\rho}^\nu+{L_z}^{\mu\rho}{{L}_\rho}^\nu
\mp\frac{1}{R^2}(L^{\mu}{L_z}^{\nu}-L^{\nu}{L_z}^{\mu})=0
\label{MCp1} 
\eea
that together with \bref{MCp} define a free differential algebra.

If we introduce the generators $Z_\mu$ and $Z_{\mu\nu}$  dual to the forms
$L^{\mu}_z$ and $ L^{\mu\nu}_z$ we have the following commutation relations
\bea
\left[   P_{a}, ~ M_{\nu b}\right]&=&i\h_{ab}\;Z_\nu,
\qquad
\left[   M_{\mu a}, ~  M_{\nu b}\right]\=i\h_{ab}\;Z_{\mu\nu}
\label{contra4}
\eea
and 
\bea
\left[   P_\mu,~  Z_{\nu}\right]&=&\pm i \frac{1}{R^2}~Z_{\mu\nu},\qquad
\qquad Z_{\mu\nu}\equiv -Z_{\nu\mu}
\nn\\
\left[   P_{\mu}, Z_{\rho\s}\right]&=&-i\h_{\mu[\rho}Z_{\s]},
\label{contra5}\\
\left[  Z_{\mu},  M_{\rho\s}  \right]&=&-i\h_{\mu[\rho}Z_{\s]},
\qquad
\left[  Z_{\mu\nu},  M_{\rho\s}  \right]\=-i\h_{\nu[\rho}Z_{\mu\s]}
+i\h_{\mu[\rho}Z_{\nu\s]}.
\label{contra6}
\eea
Notice we have a non-central extension of the original Newton-Hooke algebra,
in particular $Z$'s do not commute with $P_\mu$, see eq. \bref{contra5}.
$L^{\mu}_z$ and $L^{\mu\nu}_z$ are left invariant \MC one forms of the 
above algebra if we introduce the group parameters $c^\mu$ and $c^{\mu\nu}$
with the generators $Z_{\mu}$ and $Z_{\mu\nu}$. They are
\bea
L^{\mu}_z&=& dc^{\mu}+w^{\mu\nu}c_\nu
+e^\nu{c_\nu}^\mu-{v^\mu}_ady^a -\frac12e^\nu{v_{\nu a}}v^{\mu a}
\mp \frac12e^\mu\frac{y^ay_a}{R^2} ,
\nn\\
L^{\mu\nu}_z&=& dc^{\mu\nu} +w^{[\mu\rho}{c_{\rho}}^{\nu]}
\mp\frac{e^{[\mu}c^{\nu]}}{2R^2} \pm\frac12e^{[\mu}{v^{\nu]
a}}\frac{y_a}{R^2} +\frac12{v^{[\mu}}_adv^{\nu]a}. \label{o-}
\label{Lmunu}\eea
%\footnote
{At the level of the coset this
implies to consider
\be
\label{p3} g=g_{0}\;e^{i y^{a}P_{a}}e^{iv^{\mu a}M_{\mu a}}
\;e^{ic^{\mu}Z_{\mu}} e^{\frac{i}{2}c^{\mu\nu}Z_{\mu\nu}}.\qquad
\ee
}
The non-central algebra acts in a natural way in
a bosonic "super" space $\{x^\mu, y^a, v^{\mu a}, c^\mu, c^{\mu\nu}\}$.

In section \ref{sectBrane} we have considered the unextended NH algebra
and
we have constructed an invariant closed  $p+2$ form $\Omega_{p+2}$,
\bref{Wp+2} from $L$'s.
The potential $\Omega_{p+1}$ in \bref{Wp+1} was not expressed in terms of
$L$'s and was cohomologically non-trivial. The resulting Lagrangian was
WZ Lagrangian pseudo invariant under the NH.

In the extended algebra  instead we can construct an invariant
p+1 form  as
\be
\Omega'_{p+1}\=\frac{\ep_{\mu_0\mu_1,...,\mu_p}}{p!}L^{\mu_0}...L^{\mu_{p-1}}
L^{\mu_{p}}_z.
\label{Wp+1p}\ee
$\Omega'_{p+1}$ is an invariant p+1 form and satisfies
\be
\Omega_{p+2}=d\Omega'_{p+1}
\label{Wp+2p}
\ee
where $\Omega_{p+2}$ is given in \bref{Wp+2}. The $\Omega_{p+2}$ is 
cohomologically trivial in the extended algebra. $\Omega'_{p+1}$ differs 
from $\Omega_{p+1}$ in \bref{Wp+1} by a locally exact form
\be
\Omega'_{p+1}\=\Omega_{p+1}+d\left[
\frac{(-1)^p}{p!}\ep_{\mu_0,...,\mu_p}L^{\mu_0}...L^{\mu_{p-1}}
\;c^{\mu_p}\right].
\label{Wp+1rel}
\ee
The Lagrangian is the pullback of  $\Omega'_{p+1}$ which is the one 
associated with $\Omega_{p+1}$ plus a surface term proportional to
$c^\mu$ and it is invariant  under the extended NH algebra due to 
appropriate transformations of $c^\mu$.
%%%%%%%%%%%%%%%%%%%%%%%%%%%%%%%%%%%%%%%%%%%%%%%%%%%%%%%%%%%%%%%

\section{Generalized pp-wave metric and extended NH algebra}

Here we  construct  a non-degenerate relativistic metric from the
generators of  extended NH algebra by considering the dual forms
\bref{W1} \footnote{See the appendix for a general discussion}

Let us first consider the case of the $NH_{\pm}$ particle case $(p=0)$.
We consider the quadratic $H$ invariant combinations of the generators
\be
 C=- 2Z_0P_0 +P_a^2 + \frac{1}{R^2} M_{0 a}^2.
\ee
The associated invariant metric is given by
\be
ds^2=\left(-2 dx^0dc^0 \pm {(dx^0)}^2\frac {(y_a)^2}{R^2}+ (dy_a)^2\right)
+\left(d(R v^{0a})\mp dx^0 \frac {y^a}R\right)^2. \label{metric2}
\ee

It is the metric of a pp-wave with angular momenta,
the gyraton\footnote{We acknowledge Eloy Beato for discussions on this
point}\cite{Frolov:2005zq}\cite{Coley:2004hu}, 
note that $ x^0, c^0$ are light-like
coordinates and $\partial_c$ is a covariantly constant null Killing
vector.

The ordinary pp-wave metric is obtained by considering the first term of
\bref{metric2}. The relation among the pp-wave metric and the particle 
NH algebra was studied in \cite{Gibbons:2003rv}\footnote{
Previous discussion along this direction is found in 
\cite{Duval:1990hj}}.
This result agrees with the fact that the pp-wave algebra in $d+1$ dimensions
is isomorphic to a central extended NH algebra in d dimensions.
Note that we have obtained a  relativistic pp-wave
metric from an extended non-relativistic algebra.

In the case of the NH brane algebra the generalization of the pp-wave metric is
obtained by considering the quadratic $H$ invariant combination of the
generators 
\be
\label{casimirnh}
C=2\eta_{\mu\nu}P^\mu Z^\nu+P^aP^a
\ee
the metric is
\be
\label{metric3}
ds^2= 2L_\mu L^{\mu}_z+ L^aL^a.
\ee
Using \bref{W1} and \bref{Lmunu} we have
\be\label{nullnh}
ds^2=2 e_\mu (Dc)^\mu \mp
e_\mu e^\mu\frac {(y_a)^2}{R^2}+(dy_a)^2,
\ee
where $e^\mu, \omega^{\nu\rho}$ are $AdS_{p+1}$ vielbein and spin connection,
$(Dc)^\nu= dc^\nu +\omega^{\nu\rho}c_\rho$ is the covariant derivative.
Note that this metric contains $p+1$ covariantly constant null vectors
$\partial_{c^\mu}$, therefore this metric lives in a space with more than 
one times.

%%%%%%%%%%%%%%%%%%%%%%%%%%%%%%%%%%%%%%%%%%%%%%%%%%%%%%%%%
\section{Extended NH algebras as subalgebras of $SO(d+1,p+2)$}

The ordinary particle Galilei group
in $d$ dimensions and its  central extension,
the Bargmann algebra,
are a subgroup of Poincar\'e group in $d+1$  dimensions.
\be
 ({\rm Bargmann})_d \quad\subset\quad
({\rm Poincare})_{d+1}.
\ee
To see this fact we
introduce the light cone indices $\pm$
\be
A_\pm\equiv \frac{1}{\sqrt 2}(A_d\pm A_0).
\ee
Their generators are identified with
those of Poincare generators in $d+1$  dimension as
\be
H=P_-,\quad K_a=M_{+a},\quad  Z=P_+,\quad  M_{ab},\quad  (a=1,2,...d-1),
\ee
where $H$ and $K_a$ are energy and boost generators in the Galilei algebra
and $Z$ is the central charge.

For the particle NH algebras an analogous construction can
be done where the role of the Poincare group is taken by the
conformal group $SO(d+1,2)$. The group is linearly realized in $d+3$
dimensional space with metric $\eta_{MN}=(-;+...+,-), 
(M,N=\{0,1,...,d,d+1,d+2\})$. It is useful to introduce two light-like indices
\be
A_\pm\equiv \frac{1}{\sqrt 2}(A_d\pm A_0),\qquad
A_{\pm'}\equiv \frac{1}{\sqrt 2}(A_{d+1}\pm A_{d+2})
\ee
so that the non-zero components of the metric are
\be
\h_{+-}=\h_{+'-'}=1,\qquad \h_{ab}=\D_{ab},\qquad (a,b=1,...,d-1).
\ee
The NH algebras, NH$_\pm$, are subalgebras of $so(d+1,2)$
whose generators are expressed in terms of  $so(d+1,2)$ generators
$\CM_{MN}$ as
\bea
{P}_0&=&\frac1{R}(\CM_{-'+}\pm \CM_{+'-}),\qquad
{P}_a\=\frac1{R}\CM_{a+},\nonumber\\
{K}_a&=&\CM_{+'a},\qquad
M_{ab}\=\CM_{ab},
\nonumber\\
Z&=&\frac1{R}\CM_{+'+}.
\eea
\vs

In the case of the NH$_\pm$ algebras for general $p$-branes
we consider the "multi-temporal" conformal group in $d+1+p$ dimensions,
i.e., $SO(d+1,p+2)$. We
introduce $p+2$ sets of light-like vectors
$(\pm,\pm_0,\cdots,\pm_{p})$ with $\h_{+-}=
\h_{+_{\hskip-0.5mm \mu}-_{\hskip-0.5mm\mu}}=1, (\mu=
0,...,p)$.
The brane NH$_\pm$ generators satisfying
extended algebra \bref{contra1}-\bref{contra2} and
\bref{contra4}-\bref{contra6} are given by
\bea
{P}_\mu&=&\frac1{\7R}(\CM_{-_{\hskip-0.5mm \mu}+}\pm
\CM_{+_{\hskip-0.5mm \mu}-}),\qquad
{P}_a\=\frac1{\7R}\CM_{a+}
\nonumber\\
{M}_{\mu\nu}&=&\CM_{+_{\hskip-0.5mm \mu}-_{\hskip-0.5mm \nu}}
-\CM_{+_{\hskip-0.5mm \nu}-_{\hskip-0.5mm \mu}},\qquad
{M}_{\mu a}=\CM_{+_{\hskip-0.5mm \mu}a},\qquad M_{ab}\=\CM_{ab},
\nonumber\\
Z_\mu&=&\frac1{\7R}\CM_{+_{\hskip-0.5mm \mu}+},\qquad
Z_{\mu\nu}\=\CM_{+_{\hskip-0.5mm \mu}+_{\hskip-0.5mm \nu}}.
\eea

Summing up we have shown that the extended NH
groups for p-branes in $d$ dimensions are subgroups of multi-temporal
relativistic conformal groups in $d+p+1$ dimensions ,
\be
({\rm extended\; NH})_d \quad\subset\quad
(SO(d+1,p+2))_{d+p+1}.
\ee
%%%%%%%%%%%%%%%%%%%%%%%%%%%%%%%%%%%%%%%%%%%%%%%%%%%%

\section{Summary and Discussions}
We have constructed $NH_{\pm}$ p-brane algebras in $d$
dimensions as contractions of dS(AdS)
algebras. Non-relativistic brane actions  are constructed as WZ terms
of these algebras since the $p+2$ Eilenberg-Chevalley cohomology group
is non-trivial. These bosonic algebras have non-central extensions due to the 
existence of
a non-trivial vector valued 2 form on a representation of the
stability group. These algebras appear also as subalgebras of a
multitemporal relativistic conformal algebras, $ SO(d+1,p+2)$.

Finally we have constructed a generalization of the pp-wave
metric from quadratic $H$ invariant combinations of the generators.
For the case of NH brane algebras we have seen that these metrics
lives  in spaces with more than one time.
Since extended NH algebras appear as an special limit of string theory,
it will be interesting to see if one could give some physical
meaning to these space with more than one time.
It will be also interesting to see the relation with the two
time physics, see for example \cite{Bars:1998cs}.
\vs

\section*{Acknowledgements}

We acknowledge useful discussions with Eloy Beato, Glenn Barnich, 
Thom Curtright, Jaume Gomis, Marc Henneaux, Luca Mezincescu,
Filippo Passerini, Tonie Van Proeyen, Jorge Russo,
Paul Townsend, Peter West and Jorge Zanelli.
Joaquim Gomis and Kiyoshi Kamimura acknowledge the Perimeter Institute for the
hospitality, where a part of this work has been completed.
This work is supported in part
by MCYT FA 2004-04582-C02-01, CIRIT 2005 SGR 00564 and MRTN-CT 2004-005104.

%%%%%%%%%%%%%%%%%%%%%%%%%%%%%%%%%%%%%%%%%%%%%%%%%%%%%%%%%%%%%%%%%%%%
\appendix
\section{ Metrics from non-linear realizations}

Let us consider a  space time group $G$ with an unbroken subgroup $H$.
We split the Lie algebra generators ${\cal G}$ into $G_I\in {\cal G}-\CH$ and 
$G_i\in \CH$, where $G_I$ is generator ${\cal G}$ which does not belong to 
the stability group $H$. We consider the coset $G/H$ with following 
parametrization
\be
g=e^{iG_{I}x^{I}}
\ee
where $x^{I}$ represent all the Goldstone fields and $G_I$ all the broken 
generators.

The generators $G_{\8I}$ of the whole group $G$ transform under the action of 
${G}$ as the adjoint representation of the group
\be
G_{\8I}\to G'_{\8I}=e^{-i\ep^{\8K}G_{\8K}} G_{\8I}e^{i\ep^{\8K}G_{\8K}}=
{\Lambda_{\8I}}^{\8J}(\ep)G_{\8J},\qquad
{\Lambda_{\8I}}^{\8J}(\ep)=e^{\ep^{\8K}{f_{\8K\8I}}^{\8J}};\;
\ee
where $\8I=(I,i)$, $\ep^{\8K}$'s are the transformation parameters,
${f_{\8K\8I}}^{\8J}$ are the structure constants.

The transformation of a coset element is
\be
g\too g'\=g_0\;g\;h^{-1},\qquad g_0\in G,\qquad h\in H.
\ee
where
$h(g_0,x)$ is a compensating $H$ transformation and depends on $x$ as well 
as $g_0$ generally. The Maurer-Cartan one form transforms as
\be
\W=-ig^{-1}dg\too \Omega'=
% -i(g_0\;g\;h^{-1})^{-1}d(g_0\;g\;h^{-1})\=
h\;\W\;h^{-1}-ih\;dh^{-1}.
\ee
which implies
\bea
\Omega'_{\frac GH}&=&h\Omega_{\frac GH}h^{-1}
\nn\\
\Omega'_H&=&h\Omega_H h^{-1}-ih\;dh^{-1}.
\eea
Taking into account that
\bea
\W=-ig^{-1}dg=G_IL^I+G_iL^i
\eea
we have
\bea
{L'}^J&=&  L^I {\Lambda_{I}}^{J}(\A).
\nn\\
{L'}^i&=& L^I{\Lambda_{I}}^{i}(\A)+L^j{\Lambda_{j}}^{i}(\A) +(-ihdh^{-1})^i
\eea
where $\A\in H$ is the induced $H$  transformation  and in
general depends also on $g$.
\vs

Suppose we have a quadratic generator invariant under $H$
constructed from $G_I$'s
\be
C=g^{IJ}G_IG_J.% , \qquad  P_I\in G-H
\label{cometric}
\ee
The invariance of C under $H$ means
\be
g^{IJ}{\Lambda_{I}}^{K}(\A){\Lambda_{J}}^{L}(\A)=g^{KL},
\label{invmet}
\ee
and
\be
g^{IJ}{\Lambda_{I}}^{k}(\A){\Lambda_{J}}^{L}(\A)=0,\quad
g^{IJ}{\Lambda_{I}}^{k}(\A){\Lambda_{J}}^{\l}(\A)=0
\ee
for any $H$ transformation $\A^{i}$. If $g^{IJ}$ is not singular the first 
of above equation means ${\Lambda_{I}}^{k}(\A)=0$ and $H$ is the 
automorphism group. Note $C$ is not necessarily a Casimir
operator of ${\cal G}$. When the $g^{IJ}$ in the \bref{cometric} is not 
singular (non-degenerate) we can construct  ${G}$  invariant metric 
using its inverse; $g_{IJ}$ %g^{JK}={\D_I}^K$ as
\be
ds^2\=g_{IJ}L^IL^J.
\label{metricGH}
\ee
Actually it is invariant as
%\bea
% g_{IJ}L^IL^J \too {L'}^Tg_{..} L'\= {L}^T\Lambda g_{..}\Lambda ^T L\=L^T  
%g_{..} L.
%\eea
\bea
 g_{IJ}L^IL^J  \too g_{IJ}{L'}^I L'^J=g_{IJ}{L}^K{\Lambda_{K}}^{I}
{L}^L{\Lambda_{L}}^{J}=  g_{IJ}L^IL^J .
\eea
%Here all matrix indices are $I,J,...\in {\cal G}-\CH$. 
%In the last step we have used \bref{invmet} , which is expressed as
where we have used
\be
g^{KL}{\Lambda_{K}}^{I}{\Lambda_{L}}^{J}= 
g^{IJ} \qquad\to\qquad  g_{KL}{\Lambda_{I}}^{K}{\Lambda_{J}}^{L}=g_{IJ}.
\ee

\vs

\end{document}